\documentclass{PoS}
\pdfoutput=1

\usepackage{amsmath}
\usepackage{amsfonts}
\usepackage{bbold}
\usepackage{epsfig}

\def\l{\label}
\def\La{\mathcal{L}}

\def\({\left(}
\def\){\right)}
\def\f{\frac}
\def\be{\begin{equation}}
\def\ee{\end{equation}}
\def\bry{\begin{array}}
\def\ery{\end{array}}
\def\bes{\begin{subequations}}
\def\ees{\end{subequations}}
\def\bit{\begin{itemize}}
\def\eit{\end{itemize}}
\def\ben{\begin{enumerate}}
\def\een{\end{enumerate}}

\newcommand{\Dsl}{D\llap{/\kern+1.5pt}}
\newcommand{\MET}{E\llap{/\kern1.5pt}_T}

\title{An isosinglet $W'$ at the LHC: updated bounds from direct searches}

\ShortTitle{An isosinglet $W'$ at the LHC: updated bounds from direct searches}

\author{\speaker{Riccardo Torre}\\
        Instituto de F\'isica Te\'orica UAM/CSIC\\
        Universidad Aut\'onoma de Madrid, Cantoblanco, E-28049, Spain\\
        E-mail: \email{riccardo.torre@csic.es}}

\abstract{We study the implications of the latest LHC searches for resonances in the di-jet and di-boson final states and for quark compositeness in the di-jet angular distributions on an iso-singlet $W'$ with unit hypercharge. Predictions presented in previous papers are collected and updated to take into account all the present bounds from direct searches. It turns out that bounds on the coupling to quarks coming from the LHC measurements of di-jet angular distributions are starting to be competitive with the ones coming from direct searches in the di-jet invariant mass spectrum. Moreover, the stringent bounds on the mixing angle coming from the new searches in the $WZ\to l^{+}l^{-}l\nu$ final state, imply, in the case of discovery, the necessity to go to the LHC with \mbox{$\sqrt{s}=14$ TeV} to be able to probe interesting masses and couplings relevant for the $W'$ decay into the $W\gamma$ final state, particularly relevant to get more insight on the nature of the new vector.}

\FullConference{Proceedings of the Corfu Summer Institute 2011 School and Workshops on Elementary Particle Physics and Gravity\\
		September 4-18, 2011\\
		Corfu, Greece
		{\vspace{-20cm}\hspace{3cm}\flushright{\small \textnormal{IFT-UAM/CSIC-12-35\\
LPN12-046\\} }\vspace{20cm}}}
		
\begin{document}

\section*{}

\vspace{9cm}
\begin{flushright}
To A$\nu\tau\acute\omega\nu\eta\varsigma$
\end{flushright}

\newpage
\section{Introduction}
The Standard Model (SM) of particle physics successfully explains most of the properties and interactions of the observed particles. However, it still suffers from some theoretical issues summarized by the lack of a theoretical explanation or a complete understanding of the origin of many phenomena, among which we can mention the ElectroWeak Symmetry Breaking (EWSB) or the Higgs mechanism, the large hierarchy between different physical scales, e.g. the Fermi scale (of ElectroWeak (EW) interactions) and the Planck scale (of gravitational interactions), the generation of fermion masses and the explanation of the hierarchy between the masses of the observed fermions, the neutrino masses, the Dark Matter, the large scale structure of the observed Universe, the gravitational interactions and so on. These phenomena have been among the most relevant topics in physics for the last century leading to `giant leaps for mankind' towards answering many of the epistemological questions which are at the origin of physics. However, still a lot of work has to be done to give a final answer to most of the questions related to the aforementioned phenomena. Moreover, the lack of the possibility to directly probe energies very close to the Planck scale, where we expect the physics laws to be different form the ones we know, has required the physics community to go beyond just the mere particle physics, merging different branches of physics, like high energy physics, particle physics, cosmology, astrophysics, condensed matter physics and so on to access information on very high energy scales close to the energy scale of the Early Universe. Nowadays, impressive experimental setups allow us to access different kinds of information that can be collected to give indications of new physics, i.e.~physics Beyond the Standard Model (BSM). Among these are (have been) collider experiments like the LEP, the Tevatron and the LHC whose primary task is (has been) the discovery of the missing brick of the Standard Model, i.e. the Higgs boson.  
The recent results of the two largest experiments at the LHC, ATLAS \cite{ATLASCollaboration:2012uh} and CMS \cite{CMSCollaboration:2012tb}, as well as new updated analyses of the two Tevatron collaborations, CDF and D0 \cite{Group:2012tv}, are pointing in the direction of a relatively light SM-like Higgs boson with a mass around \mbox{$m_{h}\sim 125$ GeV}. However, in spite of the absolutely leading role of the Higgs boson searches in the current experimental strategies, many other analyses looking for new physics signals are being made. Examples are searches for Supersymmetry, Extra dimensions and new forces.\\
\indent All the new physics models that have been constructed to extend the Standard Model account for new degrees of freedom. In many of them, new spin-1 resonances are expected to show up at the TeV scale and could represent an important probe either of new gauge interactions or of a strong sector just above the Fermi scale. In the first case, while new neutral vectors, referred as $Z'$ \cite{Langacker:2008yv,Salvioni:2009p068,Salvioni:2010p010,Accomando:2010fz} are predicted already by abelian extensions of the SM gauge group $G_{\mathrm{SM}}=SU(3)_{C}\times SU(2)_{L}\times U(1)_{Y}$ in which an extra $U(1)$ gauge symmetry is considered, electrically charged vectors ($W'$) require in general non-abelian extensions of the SM gauge symmetry. Examples of these models, often inspired by grand unified theories, include Left-Right (LR) models and Little Higgs models.
In the second case, the Higgs boson and new heavy vectors can arise as composite states bound by a new strong interaction that can be responsible for the spontaneous breaking of the EW symmetry. In this case the new vectors can also play a leading role in the partial UV completion of the model up to the cut-off fixed by the strong coupling regime \cite{Csaki:2003dt,Barbieri:2008p1580,Contino:2011np}. An important example of this class are models where the Higgs boson arises as a pseudo-Goldstone boson of a global spontaneous symmetry breaking (see, e.g., Refs.~\cite{Agashe:2004ib,Agashe:2006p363,Gripaios:2009pe}). The LHC phenomenology of these composite states in Higgsless and composite Higgs models has been discussed, e.g., in Refs.~\cite{He:2008p3365, Barbieri:2010p144, Barbieri:2010p1577, CarcamoHernandez:2010p1578, Cata:2009p713,Birkedal:2004au,Martin:2009gi}. \\
\indent A bottom up approach in the study of the LHC reach on new `composite' resonances is based on general effective Lagrangians, sometimes referred to as phenomenological Lagrangians, describing the interactions of the new states with the SM particles only requiring invariance under $G_{\mathrm{SM}}$ (see, e.g., Refs.~\cite{Bauer:2010p280,Barbieri:2011p2759,Han:2010p2696,Grojean:2011vu,Pappadopulo:2011wt,Petersson:2012tl}).
Here we focus on a $W'$ transforming as a $(\mathbf{1},\mathbf{1})_{1}$ representation of $G_{\mathrm{SM}}$. This state has been shown to be weakly constrained due to its lepto-phobic nature and has been studied within an effective approach in Refs.~\cite{Grojean:2011vu} (updated in Ref.~\cite{Torre:2011vn}). In the present paper we review and update the constraints presented in those papers in the light of the 2011 LHC run using all the updated analyses available to date for the relevant final states.\\
This contribution is organized as follows. In Section \ref{section:model} we write down the phenomenological Lagrangian describing the iso-singlet $W'$, in Section \ref{section:indirectconstraints} we summarize the indirect bounds on the new charged vector, in Sections \ref{section:newconstraints} we discuss the new constraints coming from recent ATLAS and CMS analyses and in Section \ref{section:conclusion} we conclude.

\section{The phenomenological Lagrangian for an iso-singlet $W'$}\l{section:model}
At renormalizable level, the most general Lagrangian coupling a charged iso-singlet vector $V_{\mu}$ to the Standard Model fields is given by
\begin{equation}  \label{lagrangian}
\mathcal{L}=\mathcal{L}_{\text{SM}}+\mathcal{L}_{V}+\mathcal{L}_{V-\text{SM}}\,,
\end{equation}
where $\mathcal{L}_{SM}$ is the SM Lagrangian, and
\begin{align}
\mathcal{L}_{V}\,=\,& \nonumber D_{\mu}V_{\nu}^{-}D^{\nu}V^{+\mu}-D_{\mu}V_{\nu}^{-}D^{\mu}V^{+\nu}+\tilde{M}^{2}V^{+\mu}V_{\mu}^{-}\\ 
&+\frac{g_{4}^{2}}{2}|H|^{2}V^{+\mu}V_{\mu}^{-}-ig_{B}B^{\mu\nu}V^{+}_{\mu}V^{-}_{\nu}\,,\\ \label{V-SM}
\mathcal{L}_{V-SM}\,=\,& V^{+\mu}\left(ig_{H}H^{\dagger}(D_{\mu}\tilde{H})+ \frac{g_{q}}{\sqrt{2}}(V_{R})_{ij}\overline{u_{R}^{i}}\gamma_{\mu}d_{R}^{j}\right) +\text{h.c.}\,, 
\end{align}
where we have denoted the extra state with $V^{\pm}_{\mu}$, and have defined $\tilde{H}\equiv i\sigma_{2}H^{\ast}$. The coupling of $V_{\mu}$ to left-handed fermionic currents is forbidden by gauge invariance. The covariant derivative is referred to the SM gauge group: for a generic field $X$, neglecting color, we have
\begin{equation}
D_{\mu}X=\partial_{\mu}X-igT^{a}\hat{W}^{a}_{\mu}X-ig'YB_{\mu}X\,,
\end{equation}
where $T^{a}$ are the generators of the $SU(2)_{L}$ representation where $X$ lives, and we have denoted the $SU(2)_{L}$ gauge bosons with a hat, to make explicit that they are gauge (and not mass) eigenstates. In the Lagrangian \eqref{lagrangian} we have also written the heavy vector mass term explicitly: the details of the mass generation mechanism will not affect our phenomenological study, as long as additional degrees of freedom possibly associated with such mechanism are heavy enough.
Upon EWSB the coupling $g_{H}$ generates a mass mixing between $\hat{W}_{\mu}^{\pm}$ and $V_{\mu}^{\pm}$. This mixing is rotated away by introducing mass eigenstates
\begin{equation}
\begin{pmatrix} W^{+}_{\mu} \\
W^{\prime\,+}_{\mu} \end{pmatrix} = \begin{pmatrix}
\cos\hat\theta\,\,\, & \sin\hat\theta \\
-\sin\hat\theta\,\,\, & \cos\hat\theta 
\end{pmatrix} \begin{pmatrix}
\hat{W}_{\mu}^{+} \\
V_{\mu}^{+}
\end{pmatrix}\,.
\end{equation} 
The expression for the mixing angle is then
\begin{equation}
\tan(2\hat{\theta})=\frac{2\Delta^{2}}{m^{2}_{\hat{W}}-M^{2}}\,,
\end{equation}
where 
\begin{equation}\l{parameters}
m_{\hat{W}}^{2}=\frac{g^{2}v^{2}}{4},\quad \Delta^{2}=\frac{g_{H}gv^{2}}{2\sqrt{2}},\quad M^{2}=\tilde{M}^{2}+\frac{g_{4}^{2}v^{2}}{4}\,,
\end{equation}
and $v\approx 246$ GeV is the SM Higgs \textsc{vev}.\\
We left out from the Lagrangian \eqref{lagrangian}:
\bit
\item Quartic self-interactions: the operators $V_{\mu}^{+}V^{+\mu}V_{\nu}^{-}V^{-\nu}$ and $V_{\mu}^{+}V^{-\mu}V_{\nu}^{+}V^{-\nu}$ are allowed by gauge invariance. However, they are irrelevant for our analysis and they will be neglected. Notice that a cubic self-interaction of $V_{\mu}$ is forbidden by gauge invariance.
\item Higher-Dimensional Operators (HDO): they would be suppressed with respect to renormalizable ones by the cut-off of the theory. We expect HDO to give corrections roughly of order $M^{2}_{W'}/\Lambda^{2}$ to our results: in Ref.~\cite{Grojean:2011vu} it was shown that the cut-off is always, for the region of parameters we are considering, much larger than the $W'$ mass, so that we can conservatively estimate our results to hold up to 10 percent corrections due to HDO.
\item Right-handed neutrinos: we don't make any assumption about the underlying model.
\eit
We assume that the Lagrangian \eqref{lagrangian} is written in the mass eigenstate basis for fermions and that the standard redefinition of the phases of the quark fields has already been done in $\mathcal{L}_{SM}$, thus leaving only one CP-violating phase in the Cabibbo--Kobayashi--Maskawa (CKM) mixing matrix $V_{CKM}$. The \mbox{right-handed} mixing matrix $V_{R}$ does not need to be unitary in the framework we adopt here: it is in general a complex $3\times 3$ matrix. This is a relevant difference with respect to LR models, where $V_{R}$ must be unitary, as a consequence of the gauging of $SU(2)_{R}$. We normalize $g_{q}$ in such a way that $|\det (V_{R})|=1$ (a generalization of this condition can be applied if $V_{R}$ has determinant zero).

In the mass eigenstate basis both for spin-$1/2$ and spin-1 fields, the charged current interactions for quarks read
\begin{equation}
\mathcal{L}^{q}_{cc}\,=\,W^{+}_{\mu}\,\overline{u}^{i}\left(\gamma^{\mu}v_{ij}+\gamma^{\mu}\gamma_{5}a_{ij}\right)d^{j}+
W^{\prime\,+}_{\mu}\,\overline{u}^{i}\left(\gamma^{\mu}v^{\prime}_{ij}+\gamma^{\mu}\gamma_{5}a^{\prime}_{ij}\right)d^{j}+\text{h.c.}\,,
\end{equation}
where $u^{i}, d^{j}$ are Dirac fermions, and the couplings have the expressions
\bes\l{coefficients}
\begin{align}
v_{ij}=& \frac{1}{2\sqrt{2}}\left(g_{q}\sin\hat\theta(V_{R})_{ij}+g\cos\hat\theta(V_{CKM})_{ij}\right)\,, \\
a_{ij}=& \frac{1}{2\sqrt{2}}\left(g_{q}\sin\hat\theta(V_{R})_{ij}-g\cos\hat\theta(V_{CKM})_{ij}\right)\,, \\
v^{\prime}_{ij}=& \frac{1}{2\sqrt{2}}\left(g_{q}\cos\hat\theta(V_{R})_{ij}-g\sin\hat\theta(V_{CKM})_{ij}\right)\,, \\
a^{\prime}_{ij}=& \frac{1}{2\sqrt{2}}\left(g_{q}\cos\hat\theta(V_{R})_{ij}+g\sin\hat\theta(V_{CKM})_{ij}\right)\,. 
\end{align}
\ees
We note that in general $g_{H}$ is a complex parameter (e.g., in LR models \cite{Grojean:2011vu}). However, the transformation $g_{H}\to g_{H}e^{-i\alpha}$ (with $\alpha$ an arbitrary phase) on the Lagrangian \eqref{lagrangian} only results, after diagonalization of $W-W'$ mixing, in $V_{R}\to e^{i\alpha}V_{R}$, therefore its effects are negligible for our scopes. Thus for simplicity we take $g_{H}$ to be real. The charged current interactions for leptons have the form
\begin{equation} 
\mathcal{L}^{\ell}_{cc}\,=\,W^{+}_{\mu}\cos\hat\theta\frac{g}{\sqrt{2}}\overline{\nu}^{i}_{L}\gamma^{\mu}e^{i}_{L}-W^{\prime\,+}_{\mu}\sin\hat\theta \frac{g}{\sqrt{2}}\overline{\nu}^{i}_{L}\gamma^{\mu}e^{i}_{L}\,.
\end{equation}
The trilinear couplings involving the $W'$, the $W$ and the Higgs and the $W'$ and two SM gauge bosons read
\begin{subequations}
\begin{align} \nonumber
\mathcal{L}_{W'Wh}&=\Big[-\frac{1}{2}g^{2}vh\sin\hat\theta\cos\hat\theta+\frac{g_{H}g}{\sqrt{2}}vh(\cos^{2}\hat\theta -\sin^{2}\hat\theta)+\frac{g_{4}^{2}}{2}hv\sin\hat\theta\cos\hat\theta\, \Big] \\
&\hspace{4mm}\times (W^{+\,\mu}W_{\mu}^{\prime\,-}+W^{-\,\mu}W_{\mu}^{\prime\,+})\,, \\
\label{W'Wgamma-111}
\mathcal{L}_{W'W\gamma}&=-i\,e(c_{B}+1)\sin\hat\theta \cos\hat\theta F_{\mu\nu}(W^{+\,\mu}W^{\prime\,-\,\nu}+W^{\prime\,+\,\mu}W^{-\,\nu})\,, \\
\mathcal{L}_{W'WZ}&=\, i\sin\hat\theta  \cos\hat\theta\Big[(g\cos\theta_{w}+g'\sin\theta_{w})(W^{-\,\mu}W^{\prime\,+}_{\nu\mu}+W^{\prime\, -\, \mu}W^{+}_{\nu\mu}- W^{\prime\,+\, \mu}W^{-}_{\nu\mu} \nonumber\\ \label{W'WZvertex}
&\hspace{4mm} -W^{+\,\mu} W^{\prime\,-}_{\nu\mu})Z^{\nu} -(g\cos\theta_{w}-g'\sin\theta_{w}c_{B})\left(W^{+\,\mu} W^{\prime\,-\,\nu}+ W^{\prime\,+\,\mu} W^{-\,\nu}\right) Z_{\mu\nu}\Big]\,,  
\end{align}
\end{subequations}
where $\theta_{w}$ is the weak mixing angle. 
In summary, in addition to the $W'$ mass, 4 couplings appear in our phenomenological Lagrangian: $g_{q}$, $g_{H}$ (or equivalently the mixing angle $\hat\theta$), $g_{B}$ and $g_{4}$. We find it useful to normalize $g_{B}$ to the SM hypercharge coupling, so we will refer to $c_{B}\equiv g_{B}/g'$ in what follows. In Ref.~\cite{Grojean:2011vu} it was shown that the Lagrangian \eqref{lagrangian} describes, for suitable values of the parameters, the low energy limit of a LR model.\\
The partial widths corresponding to the different two-body decays of the $W'$ can be found in the Appendix A of Ref.~\cite{Grojean:2011vu}.

\section{Summary of indirect bounds}\l{section:indirectconstraints}
Indirect bounds of different origin constrain the couplings of the $W'$:
\bit
\item $g_{q}$ is mainly constrained by $K$ and $B$ meson mixings, i.e. $\Delta F=2$ transitions. The bounds are strongly dependent on the structure of the right-handed mixing matrix $V_{R}$.
\item $\hat\theta$ is constrained by ElectroWeak Precision Tests (EWPT) (especially the $\hat{T}$ parameter) and by semileptonic $u\rightarrow d$ and $u\rightarrow s$ transitions.
\item $g_{B}$ is only weakly constrained by Trilinear Gauge Couplings (TGC) measured at LEP.
\item $g_{4}$ is essentially unconstrained and marginal in our analysis. It only indirectly affects, and in a subleading way, the partial width for the decay $W'\rightarrow Wh$.
\eit
All these bounds have been extensively discussed in Ref.~\cite{Grojean:2011vu}. Here we summarize the results.

We assume the least constrained form of the right-handed mixing matrix $V_{R}$, namely
\begin{equation} \label{RHmixing}
\left|V_{R}\right|=\mathbb{1}\,,
\end{equation}
for which the bound in the $(M_{W'},g_{q})$ plane coming from the experimental determination of the parameter $\Delta m_{K}=m_{K_{L}}-m_{K_{S}}$ reads at $90 \%$ CL \cite{Langacker:1989p2578}
\begin{equation} \label{Klimit}
M_{W'}> \frac{g_{q}}{g}\,300\,\mathrm{GeV}\,.
\end{equation} 
It was shown in Ref.~\cite{Langacker:1989p2578} that this bound still holds if each $(V_{R})_{ij}$ is varied by $\epsilon=0.01$ from its central value, so that extreme fine tuning is avoided. 
Also notice that the form of the mixing matrix \eqref{RHmixing} automatically satisfies bounds coming from $b\rightarrow s\gamma$ and from $B^{0}_{d,s}$-$\overline{B}^{0}_{d,s}$ mixing.

The mixing angle $\hat\theta$, or equivalently the parameter $g_{H}$ that appears in the Lagrangian \eqref{lagrangian}, is mainly constrained by EWPT. The main constraint comes from the negative contribution to the $\hat{T}$ parameter which, at leading order in the $v^{2}/M^{2}$ expansion reads
\begin{equation} \label{Tcontribution}
\hat{T}_{V}=-\frac{\Delta^{4}}{M^{2}m_{\hat{W}}^{2}}\,.
\end{equation}
%
A recent EW fit (including LEP2 data) performed in Ref.~\cite{Aguila:2010p1781} gives at $95\%$ CL
\begin{equation} \label{LEPbound}
\left|\frac{g_{H}}{M}\right|< 0.11\, \mathrm{TeV}^{-1}\,.
\end{equation}
This constraint can be translated into a bound on the mixing angle as a function of the $W'$ mass. We have that the bound ranges from $|\hat\theta|\lesssim 4\times 10^{-3}$ for $M_{W'}=300$ GeV to $|\hat\theta|\lesssim 5\times 10^{-4}$ for \mbox{$M_{W'}=2$ TeV}.

A different kind of bound which involves both the coupling to quarks $g_{q}$ and the mixing angle $\hat\theta$ comes from the precise low energy measurement of semileptonic $u\rightarrow d$ and $u\rightarrow s$ transitions (i.e. from the measurement of the corresponding entries of the CKM matrix). From Ref.~\cite{Buras:2010p3239} we can extract the 95$\%$ CL bound
\begin{equation}
-2\times 10^{-3} < \f{g_{q}\,\hat\theta}{g}V_{R}^{ud} < 3\times 10^{-3}\,,
\end{equation}
in the case of small CP phases.
On the other hand, such a bound is strongly relaxed if CP phases in $V_{R}$ are large: in the limit of maximal CP phases, only a milder second-order constraint survives, leading (assuming $V_{R}^{ud}\approx 1$) roughly to \cite{Langacker:1989p2578}
\begin{equation}
\left|\f{g_{q}\,\hat\theta}{g}\right| < 10^{-(2\div1)}\,.
\end{equation}

The only constraint on the parameter $c_{B}$ is coming from measurements of the trilinear gauge couplings which in our case are described by the quantities \cite{TripleGaugeCouplongsWorkingGroup:1996p2576}
\begin{equation}
\Delta g_{1}^{Z}=-\sin^{2}\hat\theta(1+\tan^{2}\theta_{w})\,,\qquad \Delta k_{\gamma}=-\sin^{2}\hat\theta(1+c_{B})\,,\qquad \lambda_{\gamma}=0\,.
\end{equation}
Using the fits to LEP2 data performed by the LEP experiments \cite{DELPHICollaboration:2010p2577,ALEPHCollaboration:2004p0726,L3Collaboration:2002p151,L3Collaboration:2004p151,OPALCollaboration:2004p463} letting $\Delta g_{1}^{Z},\Delta k_{\gamma}$ free to vary while keeping fixed $\lambda_{\gamma}=0$, we get a constraint on our parameter space in the $(c_{B},\hat\theta)$ plane. By combining this limit with the upper bound on the mixing angle $\hat\theta$ quoted before, we can in principle constrain $c_{B}$. However, since as discussed above the mixing angle is required to be very small, in practice TGC constrain only extremely weakly the value of $c_{B}$. For example, using the analysis performed by the DELPHI Collaboration \cite{DELPHICollaboration:2010p2577}, we find that even considering a very large mixing angle $|\hat\theta| \sim 10^{-1}$, the wide range $-11 < c_{B}< 20$ (i.e. $-3.9< g_{B}< 7.1$) is allowed by TGC measurements at $95\%$ CL.

\section{Direct searches: Tevatron vs new LHC constraints}\l{section:newconstraints}
Tevatron and LHC searches for resonances in the di-jet, $tb$ and di-boson final states set strong bounds on the parameters of the Lagrangian \eqref{lagrangian}.
The bounds coming from the CDF di-jet search with $1.13$ fb$^{-1}$ \cite{CDFCollaboration:2009p2710}, the CDF $tb$ search with $1.9$ fb$^{-1}$ \cite{CDFCollaboration:2009p3125} and the D0 $tb$ searches with $0.9$ \cite{D0Collaboration:2008p2709} and $2.3$ fb$^{-1}$ \cite{Abazov:2011xs} have been discussed in Ref.~\cite{Grojean:2011vu}. The bounds coming from di-jet searches have also been updated in Ref.~\cite{Torre:2011vn} using the ATLAS \cite{ATLASCollaboration:2011ww} and CMS \cite{Chatrchyan:2011ns} analyses with 1 fb$^{-1}$ of integrated luminosity. Here we summarize those results and include the bounds coming from the new CMS search for quark compositeness in the di-jet angular distribution \cite{CMSCollaboration:2012ue} and the new ATLAS and CMS searches for resonances in the $WZ$ final state respectively with 1.02 fb$^{-1}$ \cite{ATLASCollaboration:2012wm} and 4.7 fb$^{-1}$ \cite{CMSCollaboration:2012to}.

\subsection{Di-jet and $tb$ searches}
\begin{figure}[h]
\begin{center}
\includegraphics[scale=0.33]{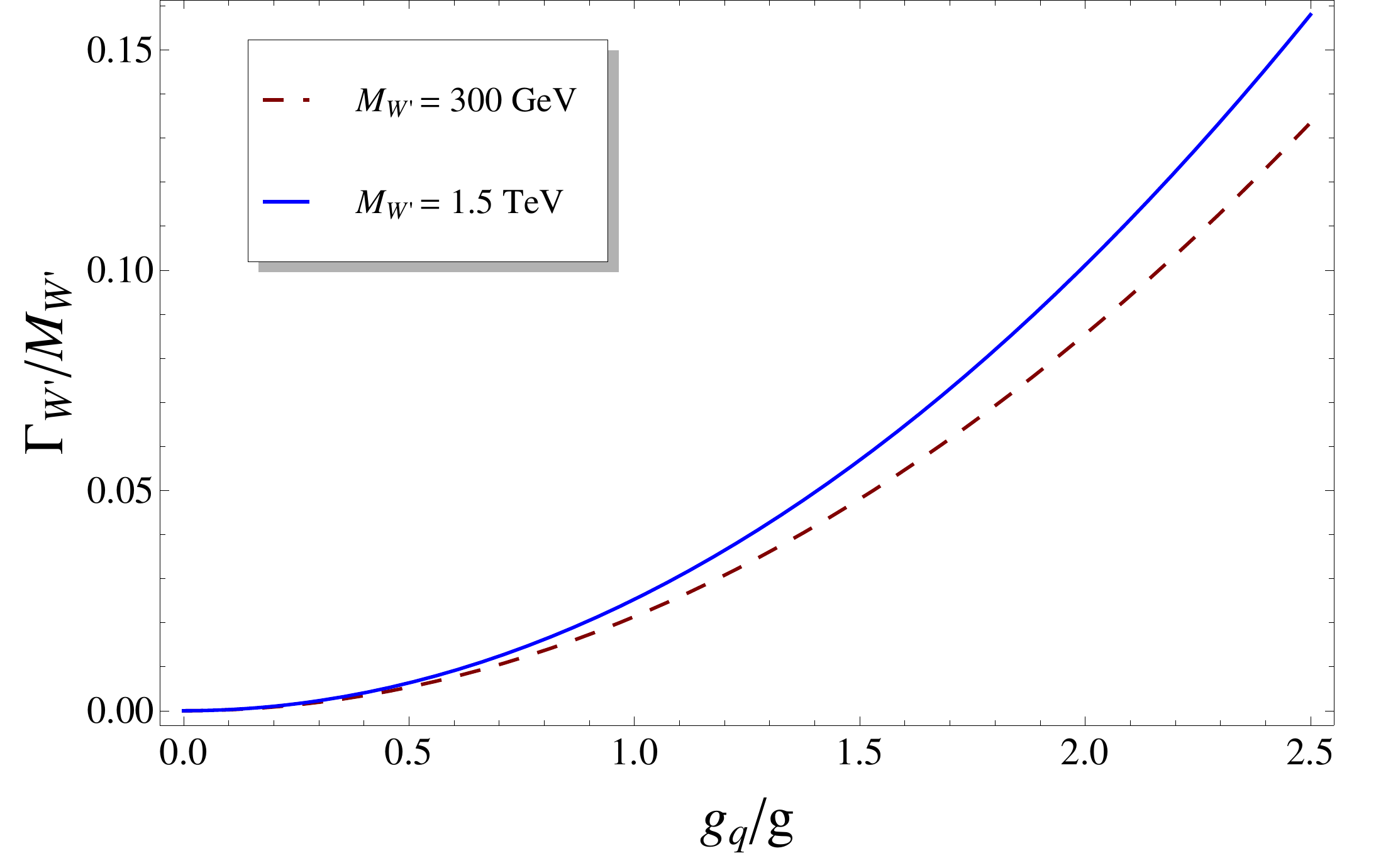}
\hspace{-1mm}\includegraphics[scale=0.33]{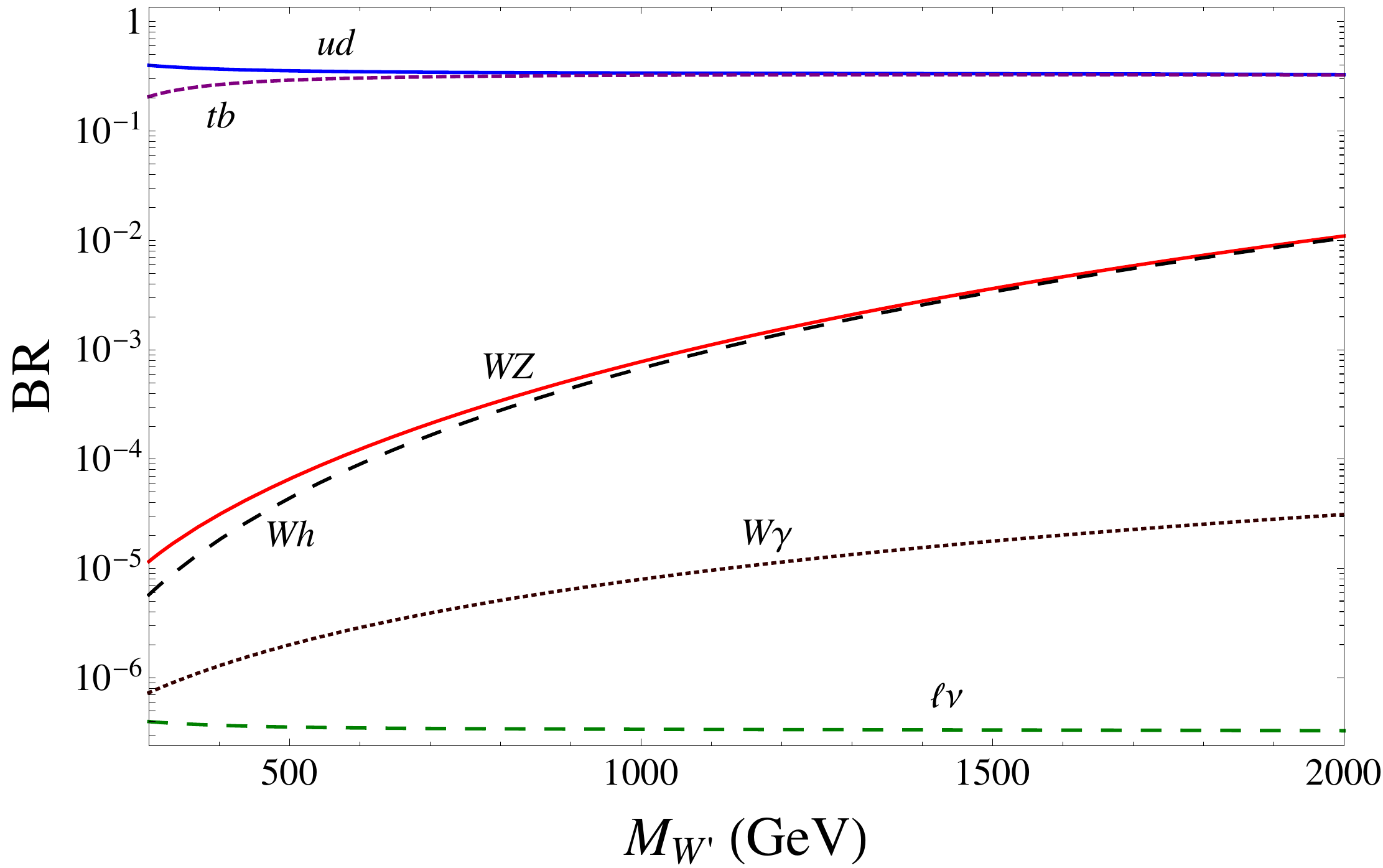}
\caption{\textsl{Left panel.} $W'$ width over mass ratio as a function of $g_{q}/g$  for negligible mixing, $\hat\theta \approx 0$, for $M_{W'}=300\,\mathrm{GeV}$ (dashed, red) and 1.5 TeV (blue). \textsl{Right panel.} BR's of the $W'$ as functions of its mass, for the following choice of the remaining parameters: $g_{q}=g$, $\hat\theta=10^{-3}$, $c_{B}=-3$, $g_{4}=g$. From top to bottom: $ud$, $tb$, $WZ$, $Wh$, $W\gamma$, $\ell\nu$ (the latter includes all the three lepton families).}
\label{fig:totalwidth}
\end{center}
\end{figure}

\l{section:dijets}
To discuss bounds on the coupling to quarks $g_{q}$, we can assume a negligible mixing angle $\hat{\theta}\approx 0$. 
In Fig.~\ref{fig:totalwidth} we show the dependence of the ratio $\Gamma_{W'}/M_{W'}$ on the normalized coupling $g_{q}/g$ (left panel) and the $W'$ decay Branching Ratios (BR's) as functions of $M_{W'}$ (right panel), for representative values of the parameters.

\begin{figure}[h!]
\begin{center}
\includegraphics[scale=0.55]{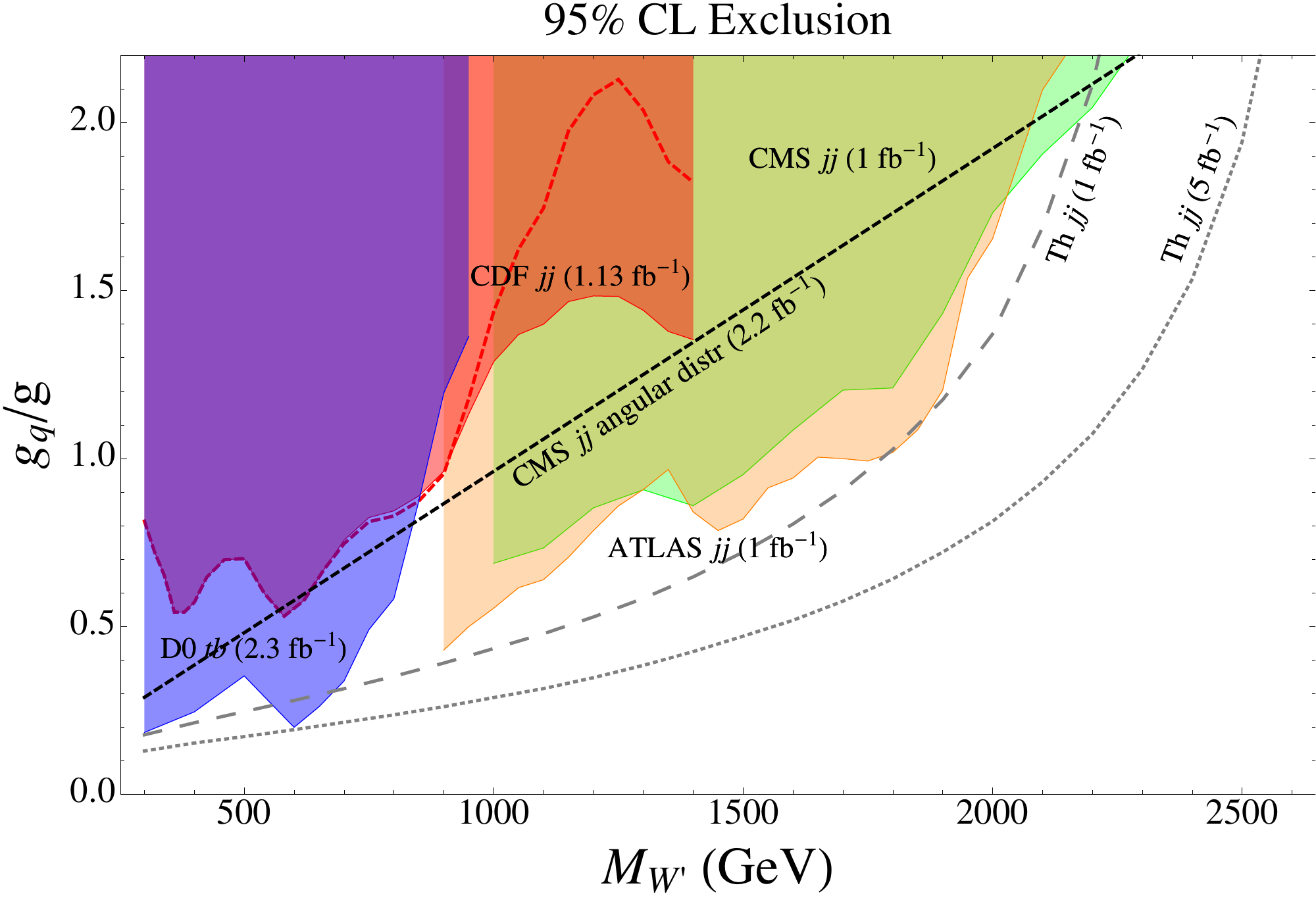}
\caption{Regions of the $\(M_{W'},g_{q}/g\)$ plane excluded at $95\%$ CL by the Tevatron and LHC searches in the di-jet invariant mass spectrum (red region $0.3-1.4$ TeV for CDF di-jet search, orange region $0.9-2.1$ TeV for ATLAS di-jet search, green region $1-2.5$ TeV for CMS di-jet search) and the D0 search in the $tb$ final state (blue region $0.3-0.95$ TeV). The black dashed line represents the bound coming from the CMS search for quark compositeness in the di-jet angular distribution. Also shown in grey are the $95\%$ CL exclusion contours computed with a parton level simulation for $1$ fb$^{-1}$ (dashed) and $5$ fb$^{-1}$ (dotted) of integrated luminosity. For further details see Ref.~\cite{Grojean:2011vu}.}
\label{fig:Bounds}
\end{center}
\end{figure}

The most recent di-jet search at the Tevatron, based on 1.13 fb$^{-1}$ of data, has been performed by the CDF Collaboration \cite{CDFCollaboration:2009p2710} while the most recent search into the $tb$ final state at the Tevatron has been performed by the D0 Collaboration and is based on 2.3 fb${}^{-1}$ of data \cite{Abazov:2011xs}. The bounds coming from these analyses on the $(M_{W'},g_{q}/g)$ parameter space have been studied in Ref.~\cite{Grojean:2011vu} and are summarized by the red ($0.3-1.4$ TeV) and the blue ($0.3-0.95$ TeV) exclusion regions in Figure \ref{fig:Bounds}, respectively for the CDF di-jet search and for the D0 $tb$ search\footnote{The D0 bound has been updated with respect to Ref.~\cite{Grojean:2011vu}.}.

Both the ATLAS and the CMS Collaborations have performed searches for resonances in the di-jet invariant mass spectrum with $1$ fb$^{-1}$ of integrated luminosity, respectively in Refs.~\cite{ATLASCollaboration:2011ww} and \cite{Chatrchyan:2011ns}. The bounds coming from these analyses are represented by the orange ($0.9-2.1$ TeV) and the green ($1-2.3$ TeV) exclusion regions in Figure \ref{fig:Bounds}. The analyses of the ATLAS and CMS Collaborations focus on high mass resonances requiring a minimum di-jet invariant mass of $0.9$ and $1$ TeV respectively. Moreover, the two limits have been set using different techniques. The CMS Collaboration has set a different limit on the production cross section for the different partonic final states $qq$, $qg$ and $gg$ by taking into account the different amount of QCD radiation generated by the different final states. On the other hand, the ATLAS Collaboration sets a model independent bound on the production cross section times the acceptance as a function of the mass and the width of the new state assuming the signal shape to be gaussian and without taking into account the different final state QCD radiation for different partonic states. The two experiments employ different kinematic requirements, summarized in Table \ref{table1}, to isolate the signal from the background and set the limit. Moreover, to take into account a readout problem in the ATLAS calorimeter in the region $-0.1<\eta_{j}<1.5$ and $-0.9<\phi_{j}<-0.5$ the acceptance for the ATLAS kinematic requirements should be reduced by a factor $0.9$ \cite{ATLASCollaboration:2011ww}\footnote{For further details on the two analyses we refer the reader to Refs.~\cite{ATLASCollaboration:2011ww} and \cite{Chatrchyan:2011ns}}.
\begin{table}[t]
\begin{center}
\begin{tabular}{c|c|c}
{\bf Variable} 		& {\bf ATLAS} 	& {\bf CMS}\\ \hline
$M_{jj}$			&  --- 			&  $> 838$ GeV\\
$p_{T_{j}}$		&  $> 180$ GeV 	&  $> 717$ GeV\\
$|\eta_{j}|$			& $< 2.8$			& $< 2.5$\\
$|\Delta \eta_{jj}|$	& $< 1.2$			& $< 1.3$\\
\hline
\end{tabular}
\end{center}
\caption{\small\em Kinematic requirements used by the ATLAS and CMS analyses. The ATLAS analysis also excluded the region $-0.1<\eta_{j}<1.5$ and $-0.9<\phi_{j}<-0.5$, affected by a temporary readout problem \cite{ATLASCollaboration:2011ww}.}\label{table1}
\end{table}
All the di-jet bounds in Figure \ref{fig:Bounds} have been set comparing the experimental limits with a parton level simulation performed using the CalcHEP matrix element generator \cite{Pukhov:1999p1261,Pukhov:2004ca,Belyaev:Df-Kj9yc}.\\
Figure \ref{fig:Bounds} shows that, in spite of the different analysis techniques, the exclusion limits of the \mbox{ATLAS} and CMS Collaborations are compatible with each other. Moreover, both these limits are in reasonable agreement with the parton level prediction obtained in Ref.~\cite{Grojean:2011vu} and represented by the dashed grey line in Figure \ref{fig:Bounds}. This shows in particular that a reasonable prediction for the exclusion/discovery of a new physics model in the di-jet invariant mass spectrum can be obtained with a simple parton level simulation, provided that the signal and background distributions are compared by integrating over $M_{jj}>M_{\text{res}}(1-\epsilon/2)$ rather than in the single bin centered on the resonance mass $M_{\text{res}}(1-\epsilon/2)<M_{jj}<M_{\text{res}}(1+\epsilon/2)$, where $\epsilon$ is the expected experimental resolution on the di-jet invariant mass. In fact, the former method is less sensitive to smearing effects generated by hadronization and jet reconstruction, which are not taken into account in a parton-level analysis \cite{Grojean:2011vu}.

A constraint of different origin in the $\(M_{W'},g_{q}\)$ plane comes from the CMS search for quark compositeness in the di-jet angular distribution with $2.2$ fb$^{-1}$ of integrated luminosity \cite{CMSCollaboration:2012ue}. To compute this constraint we consider the following four-quark interaction Lagrangian, obtained by integrating out the $W'$ (in the limit $\hat\theta\approx0$)
\be\l{leff}
\La_{\text{eff}}^{(4q)}=\frac{g_{q}^{2}}{2M_{W'}^{2}}(V_{R})_{ij}^{\dag}(V_{R})_{kl}\overline{d}^{i}_{R}\gamma^{\mu}d^{l}_{R} \overline{u}^{k}_{R}\gamma^{\mu}u^{j}_{R}+\text{h.c.}\,,
\ee
where we have made use of Fierz identities. From the CMS analysis of Ref.~\cite{CMSCollaboration:2012ue}, taking into account Ref.~\cite{Pomarol:2012tc}, we get the constraint
\begin{equation} \label{angdistlimit}
M_{W'}> \frac{g_{q}}{g}\,1.04\,\mathrm{TeV}\,.
\end{equation} 
This last constraint is represented in Figure \ref{fig:Bounds} by the black dashed line. It is worth noting that this last bound is starting to be competitive with the ones coming from the searches in the di-jet invariant mass spectrum, especially for strongly coupled and for heavy resonances. 

\subsection{Di-boson searches}
\begin{figure}[t]
\centering
\includegraphics[scale=0.55]{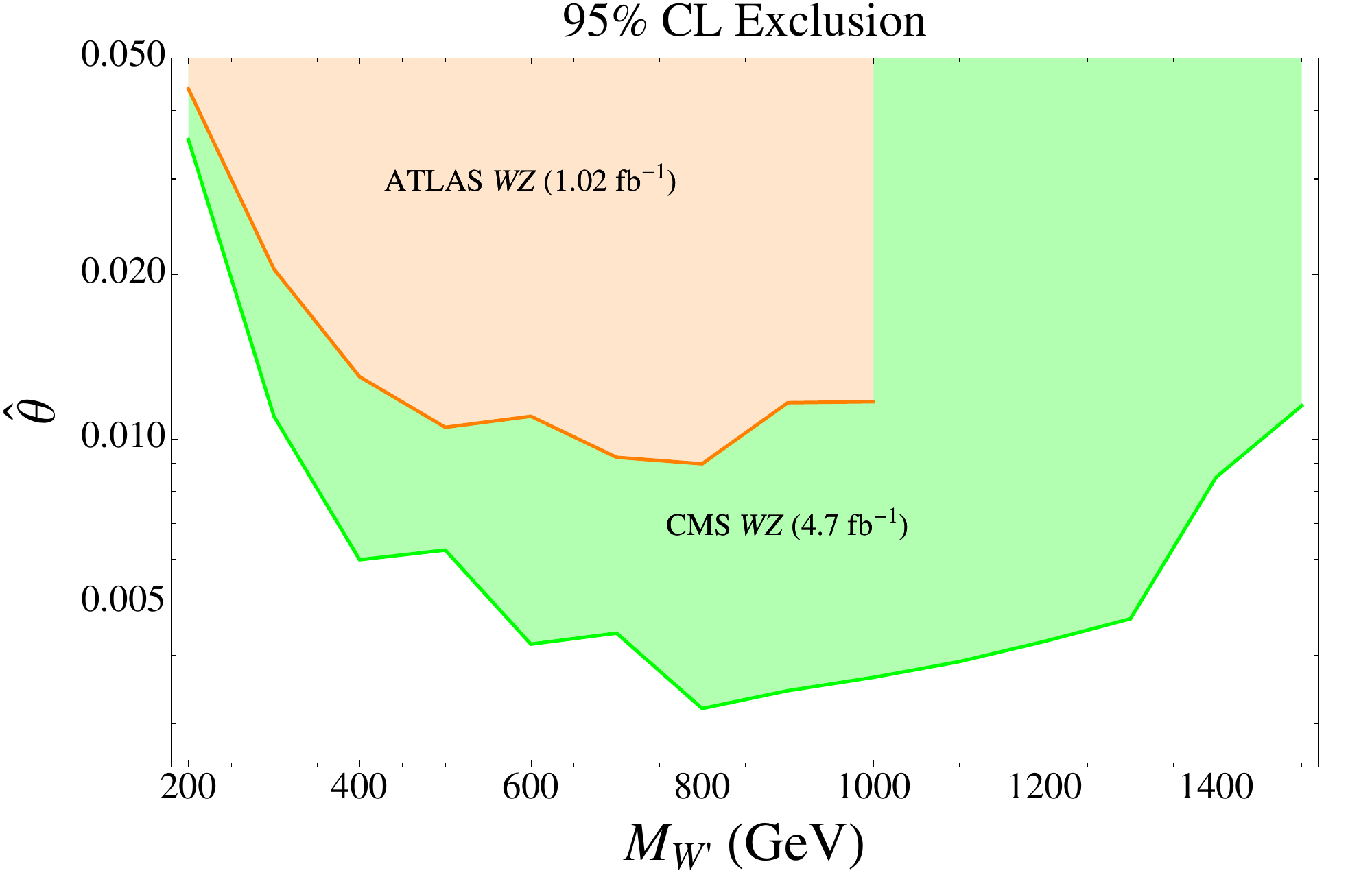}
\caption{Regions of the $\(M_{W'},\hat{\theta}\)$ plane excluded at $95\%$ CL by the LHC searches for a $W'$ in the \mbox{$WZ\to l^{+}l^{-}l\nu$} final state (orange region $0.2-1$ TeV for ATLAS, green region $0.2-1.5$ TeV for CMS).}
\label{fig:WZexclusion}
\end{figure}

\l{section:dibosons}
Both the ATLAS and the CMS Collaborations have recently performed searches for resonances in the $WZ$ final state (in the fully leptonic channel $l^{+}l^{-}l\nu$) respectively with 1.02 fb$^{-1}$ \cite{ATLASCollaboration:2012wm} and \mbox{4.7 fb$^{-1}$} \cite{CMSCollaboration:2012to} of integrated luminosity. These searches set new constraints on the $W-W'$ mixing angle $\hat{\theta}$. We have evaluated these constraints by comparing the experimental limits given in terms of an exclusion region in the $\(M_{W'},\sigma\(pp\to W'\)\times \text{BR}\(W'\to WZ\)\)$ plane with the theoretical prediction for the same quantity obtained with the CalcHEP matrix element generator. Since the production cross section depends on the $W'$ coupling to quarks $g_{q}$, the strongest bound for each value of $M_{W'}$ has been obtained choosing the corresponding upper bound on $g_{q}$ obtained in the previous Section and shown in Figure \ref{fig:Bounds}. The constraints on the mixing angle $\hat{\theta}$ coming from these analyses are shown in the $\(M_{W'},\hat{\theta}\)$ plane in Figure \ref{fig:WZexclusion}. The new LHC bound turns out to be more stringent than the old Tevatron bound shown, e.g. by the grey region in the left panel of Fig.~8 of Ref.~\cite{Grojean:2011vu}. For instance, for $M_{W'}=800$ GeV the bound changes from $\hat\theta<0.013$ (D0 \cite{D0Collaboration:2010p3231}) to $\hat\theta<0.003$ (CMS \cite{CMSCollaboration:2012to}). Also notice that the new LHC bound starts to be competitive with the constraints coming from EWPT and from flavor physics discussed in Section \ref{section:indirectconstraints}. The new direct bounds on the mixing angle $\hat{\theta}$, put strong constraints on the possibility to observe the $W'$ decay into $W\gamma$, very important to gain insight into the nature of the putative new charged vector \cite{Grojean:2011vu}.

%

\section{Conclusion}\l{section:conclusion}
In this Proceeding we have considered, in addition to the SM particle content, a new iso-singlet vector with unit hypercharge. At the renormalizable level, the most general effective Lagrangian describing the couplings of the new vector $V$ to the SM particles has been written following the approach of Ref.~\cite{Grojean:2011vu}. The main features of the model are the absence of couplings of $V$ to left-handed currents and the presence of the two operators $ig_{H}H^{\dagger}(D_{\mu}\tilde{H})V^{+\mu}$ and $ig_{B}B^{\mu\nu}V^{+}_{\mu}V^{-}_{\nu}$ which, upon EWSB, generate respectively a $V-\hat{W}$ mass mixing and a $W'W\gamma$ coupling that is forbidden in models where the new vector is a gauge boson. These features, on one hand imply that the $W'$ can be very weakly constrained by direct searches (the coupling to leptons is suppressed by the mixing angle $\hat\theta$) and flavor physics (depending on the form of the right-handed mixing matrix $V_{R}$) and on the other hand provide peculiar phenomenology in the $W\gamma$ final state. 

The indirect bounds coming from EWPT, flavor physics and TGC measurements have been summarized in Section \ref{section:indirectconstraints}, but the main goal of this contribution is contained in Section \ref{section:newconstraints} where the bounds from direct searches have been summarized and updated using the latest LHC results. The main results of this update are the following:
\bit
\item Bounds on the coupling to quarks $g_{q}$ from searches for new physics in the di-jet angular distributions start to be competitive with the ones coming from searches in the di-jet invariant mass spectrum, especially for strongly coupled and heavy resonances.
\item The new strong bounds on the $W-W'$ mixing angle $\hat\theta$ (approximately a factor of four stronger for a $W'$ mass of $800$ GeV) coming from the ATLAS and CMS searches for resonances in the $WZ\to l^{+}l^{-}l\nu$ final state disfavor the possibility, in case of discovery, to observe the decay channel $W'\to W\gamma$ at the LHC at $\sqrt{s}=8$ TeV \cite{Grojean:2011vu}.
\eit
In spite of the new stronger bounds on the production cross section ($g_{q}$) and on the BR's into $WZ$ and $W\gamma$ ($\hat\theta$), in case of discovery, the theoretical importance of the $W'\to W\gamma$ decay channel to gain insight into the nature of the new vector, motivate new searches in this channel with the full LHC center of mass energy $\sqrt{s}=14$ TeV.

\section*{Acknowledgments}
I thank C.~Grojean and E.~Salvioni for the collaboration in the original work and E.~Salvioni also for discussions and help for this update. I'm particularly grateful to the organizers of the {\it Corfu Summer Institute 2011 School and Workshops on Elementary Particle Physics and Gravity} for the invitation to submit this contribution and to the Corfu Island for the continuous inspiration. This work was partially supported by the Research Executive Agency (REA) of the European Union under the Grant Agreement number PITN-GA-2010-264564 (LHCPhenoNet).

\bibliographystyle{JHEP}
\bibliography{proceeding}

\end{document}